\documentclass[aps,preprint,showpacs,preprintnumbers,amsmath,amssymb]{revtex4}
\usepackage[dvips]{graphicx}
\usepackage{float}
\usepackage{subfigure}
\usepackage{subfig}

\begin{document}

\title{Production of scalar particles in electric field on de Sitter expanding Universe}
\author{Mihaela-Andreea B\u{a}loi}
 \email{mihaela.baloi88@e-uvt.ro}
 \affiliation{Faculty of Physics, West University of Timi\c soara,  \\V. Parvan
 Avenue 4, RO-300223 Timi\c soara,  Romania}

\begin{abstract}
The scalar particle production from vacuum in the presence of an electric field, on the de Sitter spacetime is studied. We use perturbation methods to define the transition amplitude. We obtain that the momentum is not conserved in this process. The probability density of pair production is computed by squaring the transition amplitude. Our graphical representations show that, the probability of scalar particle production was  important only in the early stages of the Universe, when Hubble's constant was very large in comparison with  the  mass of the particle. Also we propose here a criterion for particle-antiparticle separation.
\end{abstract}


\maketitle

\section{Introduction}
The theory of scalar field on de Sitter metric has been intensively studied in the literature \cite{B5,B8,B14,B17}. The solution of Klein-Gordon equation in this geometry was obtained and in addition the form of the Feynman propagator was found. But the interaction processes in scalar QED were less studied using perturbations and exact solution of field equations.
Our aim in this paper is to study the production of massive scalar particles in an external electric field on de Sitter universe. More precisely we want to demonstrate that, the process $vacuum\rightarrow \varphi+\varphi^{+}$ in the presence of an external electric field $\vec{E}$, has a nonvanishing probability only in the large expansion of early universe. Here we denote by $\varphi$, the massive scalar particle. The subject of particle creation on de Sitter expanding universe was treated by many authors \cite{B1,B2,B3}\cite{B7,B8,B9}\cite{B10,B11,B18}. Among them we mention the results of Parker, who was using the WKB approach to find the quantum mods of Klein-Gordon equation \cite{B2}. He obtained an expression for the number of particles as a function of time and showed that, the production of particles in an expanding background occurs in pairs. For our study we use exact solutions of  scalar field proposed in \cite{B5}, which have a defined momentum. We mention that our calculations are without approximations, because we use  perturbation theory to define the transition amplitude. In other words, we perform this study using the QFT formalism on Minkowski space adapted to de Sitter space \cite{B11,B12,B13}. An important result obtained in this paper is that the momentum is not conserved in this process. A similar result was obtained in \cite{B6}, where the study of fermion production from vacuum, in the presence of Coulomb field was performed. \\
Our paper is organized as follows:
in the second section we  calculate the expression of the transition amplitude. For this study we have to analyse two cases. The first case is the calculation of  the transition amplitude when  the Hankel function has a real index, which corresponds to early universe, when the expansion factor was larger than the mass of the particle. The second case is the evaluation of the transition amplitude when the  Hankel function has an imaginary index. This is the case when the mass of the particle is larger than the expansion factor and thus we can study the Minkowski limit of our results \cite{B12,B13,B21}. Also, in this section we explain why the momentum in not conserved in this process. \\
In the third  section we calculate the probability density of pair production and make a detailed analysis of the two cases mentioned above. Also, we graphically represent the probability density as a function of ratio mass/expantion factor. Our graphs show that, the probability of pair production of scalar particles in the presence of an external electric field was important in the limit of early universe. Another important thing to mention is that, we discuss here the separation between matter and antimatter. \\
In section four we present our conlusions.

\section{The transition amplitude}
In this section we present the calculation of the transition amplitude for the process of scalar pair production from vacuum in the presence of an external electric field $\vec{E}$. The de Sitter line element, in the chart $\{t,\vec{x}\}$ is \cite{B19,B22}:
\begin{equation}
ds^{2}= dt^{2}-e^{2\omega t}d\vec{x}^{2},
\end{equation}
where $\omega>0$ is the Hubble constant.
In this paper we will use quantum modes of the Klein-Gordon equation in momentum basis \cite{B5}. The solution of the scalar field on the de Sitter spacetime is correctly normalized and forms a complete set as was shown in \cite{B5}. By $f_{\vec{p}}(x)$ we denote the solution of positive frequency \cite{B5}:
\begin{equation}
f_{\vec{p}}(x)=\frac{1}{2}\sqrt{\frac{\pi}{\omega}}\frac{e^{-\pi \mu/2}}{(2\pi)^{3/2}}\,e^{-3\omega t/2}H_{i\mu}^{(1)}\left(\frac{p}{\omega}e^{-\omega t}\right)e^{i\vec{p}\cdot\vec{x}},
\end{equation}
where $H_{i\mu}^{(1)}$ is the Hankel function of the first kind and $\mu=\sqrt{\left(\frac{m}{\omega}\right)^{2}-\frac{9}{4}}$. The solution of negative frequency is $f^*_{\vec{p}}(x)$. For $t\rightarrow-\infty$, the modes defined above behave as a positive/negative frequency modes with respect to the conformal time $t_{c}=-\frac{e^{-\omega t}}{\omega}$:
\begin{equation}
f_{\vec{p}}(t,\vec{x}\,)\sim e^{-ipt_{c}};\,\,f^*_{\vec{p}}(t,\vec{x}\,)\sim e^{ipt_{c}},
\end{equation}\label{bd}
which is the behavior in the infinite past for positive/negative frequency modes which defines the Bunch-Davies vacuum.

The expression of the  transition amplitude for pair production is obtained using perturbational methods \cite{B6}, \cite{B24}, \cite{B11} and has the following form:
\begin{equation}
S_{\varphi\,\varphi^{+}}= -e\int d^{4}x \sqrt{-g}\left(f_{\vec{p}}^{*}(x)\stackrel{\leftrightarrow}{\nabla} f_{\vec{p}\,'}^{*}(x)\right)\vec{A}(x),   \label{a1}
\end{equation}
where $e$ is the coupling constant between fields and $\sqrt{-g}=e^{3\omega t}$. \\
For our study we have to find the form of the electric field on de Sitter geometry. We specify that we use the electric field produced by a point charge in this geometry. This can be done easily if we take into consideration the conformal invariance of Maxwell equations \cite{B4}. The vector potential in de Sitter geometry can be expressed in terms of the vector potential from flat space as $ A^{\mu}=\Omega^{-1}A_{M}^{\mu}$, where $\Omega =\frac{1}{(\omega t_{c})^{2}}$ \cite{B4}. If we consider $\vec{E}= -\frac{\partial\vec{A}}{\partial t}=\frac{Q}{|\vec{x}|^{2}}e^{-2\omega t}\vec{n}$, then the vector potential that produce the electric field  has the following form:
\begin{eqnarray}
\vec{A}=\frac{Q}{2\omega|\vec{x}|^{2}}e^{-2\omega t}\vec{n}\,,   \label{c1}
\end{eqnarray}
where $\vec{n}$ gives the orientation of the electric field. \\
If we look carefully at the solution of the Klein-Gordon equation, we observe that, the index of the Hankel function, $i\mu$ can also become real when $\frac{m}{\omega}<\frac{3}{2}$. This case contain the case when the expansion factor was large comparatively with the particle mass, which could be interesting if we want to address the problem of particle production in the early universe. In this case the index $\alpha$\ of the Hankel function, will become: $\alpha=\sqrt{\frac{9}{4}-(\frac{m}{\omega})^{2}}$.
Our calculations for the amplitude of the scalar particle production on the de Sitter spacetime will be done with both solutions of real and imaginary index of the Klein-Gordon equation and the results will be compared.

\subsection{The transition amplitude in the case $\frac{m}{\omega}<\frac{3}{2}$}

The relation between the two indices of the Hankel function is very simple: $\mu=i\alpha$. In this circumstances the solution of the scalar field with Hankel function being real is:
\begin{equation}
f_{\vec{p}}(x)=\frac{1}{2}\sqrt{\frac{\pi}{\omega}}\frac{e^{-i\pi\alpha/2}}{(2\pi)^{3/2}}\,e^{-3\omega t/2}H_{-\alpha}^{(1)}\left(\frac{p}{\omega}e^{-\omega t}\right)e^{i\vec{p}\cdot\vec{x}}.
\end{equation}
Using the above solutions of the Klein-Gordon equation and the vector potential given by (\ref{c1}), with $Q=e$, the expression of the transition amplitude becomes:
\begin{equation}
S_{\varphi\varphi^{+}}=\,\frac{ie^2\,(\vec{p}\,'-\vec{p}\,)\vec{n}}{64\pi^{2}\omega}\,\,e^{i\pi\alpha}\int_{-\infty}^{\infty}dt\frac{e^{-2\omega t}}{\omega}H_{-\alpha}^{(2)}\left(\frac{p}{\omega}e^{-\omega t}\right)H_{-\alpha}^{(2)}\left(\frac{p\,'}{\omega}e^{-\omega t}\right)\int_{0}^{\infty} d^{3}x\,\frac{e^{-i(\vec{p}+\vec{p}\,')\vec{x}}}{|\vec{x}|^{2}},
\end{equation}
where the volume element is $d^{4}x=d^{3}x\,dt$, and $t$ denotes the proper time.
The spatial integral is easy to solve if we consider spherical coordinates. Using the formula $(\ref{a2})$ from appendix, the expression of this integral is:
\begin{eqnarray}
\int_{0}^{\infty} d^{3}x\,\frac{e^{-i(\vec{p}+\vec{p}\,')\vec{x}}}{|\vec{x}|^{2}}= 4\pi\int_{0}^{\infty}dr \frac{\sin(|\vec{p}+\vec{p}\,'|r)}{|\vec{p}+\vec{p}\,'|r}= \frac{2\pi^{2}}{|\vec{p}+\vec{p}\,'|}.
\end{eqnarray}
If we use a new type of variable $z=\frac{e^{-\omega t}}{\omega}$\,, the temporal integral becomes:
\begin{equation}
\int_{0}^{\infty}dz z H_{-\alpha}^{(2)}(pz)H_{-\alpha}^{(2)}(p\,'z).
\end{equation}
In order to obtain the expression  of the temporal integral it is necessary to make a few steps. First, we express the Hankel function with the help of Macdonald function, using the formula $(\ref{a3})$ from appendix. Then we  use $(\ref{a4})$  to solve the new integral, and  finally, we use the relation $(\ref{a5})$ between the hypergeometric function $_{2}F_{1}$ and gamma-Euler function. The final result of the temporal integral in the case $p>p\,'$ is :
\begin{eqnarray}
&&\int_{0}^{\infty}dz z H_{-\alpha}^{(2)}(pz)H_{-\alpha}^{(2)}(p\,'z)=\,-\frac{2}{\pi^{2}}e^{-i\pi\alpha}\left[\theta(p-p\,'){\frac{1}{p^{2}}\left(\frac{p\,'}{p}\right)^{-\alpha}}
\Gamma(1-\alpha)\Gamma(\alpha) \right.\nonumber\\
&&\left.\times_{2}F_{1}\left(1,1-\alpha;1-\alpha;\left(\frac{p\,'}{p}\right)^2\right) +\theta(p-p\,')\frac{1}{p^{2}}\left(\frac{p\,'}{p}\right)^{\alpha}\Gamma(1+\alpha)\Gamma(-\alpha)\,\right.\nonumber\\
&&\left.\times_{2}F_{1}\left(1,1+\alpha;1+\alpha;\left(\frac{p\,'}{p}\right)^2\right) \right].
\end{eqnarray}
With the results of spatial, respectively temporal integrals, the final form of the transition amplitude of pair production on the de
Sitter spacetime is:
\begin{equation}
S_{\varphi\varphi^{+}}=\,-\frac{ie^2\,}{16\pi^{2}m}\frac{(\vec{p}-\vec{p}\,')\vec{n}}{|\vec{p}+\vec{p}\,'|}\left[\frac{1}{p\,'^{2}}\theta(p\,'-p)f_{\alpha}
\left(\frac{p}{p\,'}\right)+\frac{1}{p^{2}}\theta(p-p\,')f_{\alpha}\left(\frac{p\,'}{p}\right)\right],   \label{b2}
\end{equation}
where the function $f_{\alpha}\left(\frac{p\,'}{p}\right)$ was introduced to simplify the formula of the transition amplitude. The expression of this function reads:
\begin{eqnarray}\label{g2}
&&f_{\alpha}\left(\frac{p}{p\,'}\right)= k\left(\frac{p}{p\,'}\right)^{\alpha}\Gamma(1-\alpha)\Gamma(\alpha)\,_{2}F_{1}\left(1,1-\alpha;1-\alpha;\left(\frac{p}{p\,'}\right)^2\right) \nonumber \\
&&+k\left(\frac{p}{p\,'}\right)^{-\alpha}\Gamma(1+\alpha)\Gamma(-\alpha)\,_{2}F_{1}\left(1,1+\alpha;1+\alpha;\left(\frac{p}{p\,'}\right)^2\right)\nonumber\\
&&=\frac{k}{\left(1-\left(\frac{p}{p\,'}\right)^2\right)}\left[\left(\frac{p}{p\,'}\right)^{\alpha}\Gamma(1-\alpha)\Gamma(\alpha)+\left(\frac{p}{p\,'}\right)^{-\alpha}
\Gamma(1+\alpha)\Gamma(-\alpha)\right]
\end{eqnarray}
where $k=\frac{m}{\omega}$ and with the specification that $f_{\alpha}\left(\frac{p\,'}{p}\right)$ is obtained when $p\,'\leftrightarrows p$. Our result for amplitude (\ref{g2}) is proportional with $\Gamma(-\alpha)$, where $\alpha=\sqrt{\frac{9}{4}-\left(\frac{m}{\omega}\right)^{2}}$ is a real number. Because we want to approach the case of early Universe ($\omega>>m$) we will consider $m/\omega\in[0,1]$. In this interval our gamma-Euler functions are well defined and have no poles.
From eq.$(\ref{b2})$ it is easy to see that, the amplitude of pair production in the presence of an electric field is nonzero only when the momenta of the particles are not equal. As we can see from eq.$(\ref{b2})$ and eq.$(\ref{g2})$ the final expression of the transition amplitude depends on the Heaviside step function, gamma-Euler function and hypergeometric Gauss function. The hypergeometric function becomes divergent if the ratio $p\,'^{2}/p^{2}$ is $1$.  In consequence, the law of momentum conservation is not respected in this case. We know that, the de Sitter metric is invariant under spatial translations, and the momentum is conserved. So the fact that, the law of momentum conservation is broken in this process is due to the external electric field.

\subsection{The transition amplitude in the case $\frac{m}{\omega}> \frac{3}{2}$}

The transition amplitude for pair production in external electric field, for the imaginary case, can be calculated using the same methods like in the real case. A quick method to obtain the final expression of the amplitude of pair production is to make the substitution $\alpha=-\,i\mu$ in the previous result. With this replacement the transition amplitude is:
\begin{eqnarray}
S_{\varphi\,\varphi^{+}}=\frac{ie^2}{16\pi^{2}m}\frac{(\vec{p}-\vec{p}\,')\vec{n}}{|\vec{p}+\vec{p}\,'|}\left[\frac{1}{p^{2}}\theta(p-p\,')f_{\mu}\left(
\frac{p\,'}{p}\right)+\frac{1}{p\,'^{2}}\theta(p\,'-p)f_{\mu}\left(\frac{p}{p\,'}\right)\right],\label{re1}
\end{eqnarray}
where the function $f_{\mu}\left(\frac{p}{p\,'}\right)$ has the expression:
\begin{eqnarray}
&&f_{\mu}\left(\frac{p\,'}{p}\right)=k\left(\frac{p\,'}{p}\right)^{i\mu}\Gamma(1+i\mu)\Gamma(-i\mu)\,_{2}F_{1}\left(1,1+i\mu;1+i\mu;\left(\frac{p\,'}{p}\right)^2\right) \nonumber\\
&&+k\left(\frac{p\,'}{p}\right)^{-i\mu}\Gamma(1-i\mu)\Gamma(i\mu)\,_{2}F_{1}\left(1,1-i\mu;1-i\mu;\left(\frac{p\,'}{p}\right)^2\right)\nonumber\\
&&=\frac{k}{\left(1-\left(\frac{p\,'}{p}\right)^2\right)}\left[\left(\frac{p\,'}{p}\right)^{i\mu}\Gamma(1+i\mu)\Gamma(-i\mu)+
\left(\frac{p\,'}{p}\right)^{-i\mu}\Gamma(1-i\mu)\Gamma(i\mu)\right]
\end{eqnarray}

In the next section, we will use this result to calculate the probability of pair production in the  limit $\frac{m}{\omega}>\frac{3}{2}$. This amplitude corresponds to the case when the particle mass is larger than the Hubble constant.

\section{The probability of pair production}
Because, in the previous section, we discussed two cases of the transition amplitude, one corresponding to a situation with large expansion factor $(\frac{m}{\omega}<\frac{3}{2})$ and one corresponding to small expansion factor in comparison with particle mass $(\frac{m}{\omega}>\frac{3}{2})$, we also have to study the corresponding probabilities of pair production for these two cases. Also, another important thing at this section, will be the graphical representation of the density probability of pair production as a function of $m/\omega$. From this graphs we want to give  a correct interpretation of our analytical results.\\
If we  square the scattering amplitude, we obtain the probability density of transition:
\begin{equation}
P_{\varphi\varphi^{+}}= |S_{\varphi\varphi^{+}}|^{2}.  \label{b1}
\end{equation}
The probability density of pair production for $\frac{m}{\omega}<\frac{3}{2}$\,, has the following form:
\begin{eqnarray}
P_{\varphi\varphi^{+}}= \frac{e^{4}}{256\pi^{2}m^{2}}\frac{((\vec{p}-\vec{p}\,')\vec{n})^{2}}{|\vec{p}+\vec{p}\,'|^{2}}\left[\frac{1}{p^{4}}\theta(p-p\,')
\left|f_{\alpha}\left(\frac{p\,'}{p}\right)\right|^{2}+\frac{1}{p\,'^{4}}\theta(p\,'-p)\left|f_{\alpha}\left(\frac{p}{p\,'}\right)\right|^{2}\right].\label{p1}
\end{eqnarray}
The total probability is obtained by integrating after the final momenta the probability formula (\ref{p1}). These integrals are complicated and we restrict to study the properties of the density of probability.
Further, we plot the probability density of scalar particle production as a function of parameter $k=m/\omega$\,, for different values of $p/p\,'$. Here $p$ and $p\,'$ represents the momenta of the particle, respectively of the antiparticle. We observe that, the  function $f_{\alpha}$, which defines the probability, depends on $\alpha=\sqrt{\frac{9}{4}-(\frac{m}{\omega})^{2}}$. So, the numerical values taken by $m/\omega$ are in the interval $[0,\,1]$, since this interval include the interesting case of large expansion factor relatively to the particle mass.

\begin{figure}[h!t]
\centerline{\includegraphics[scale=0.5]{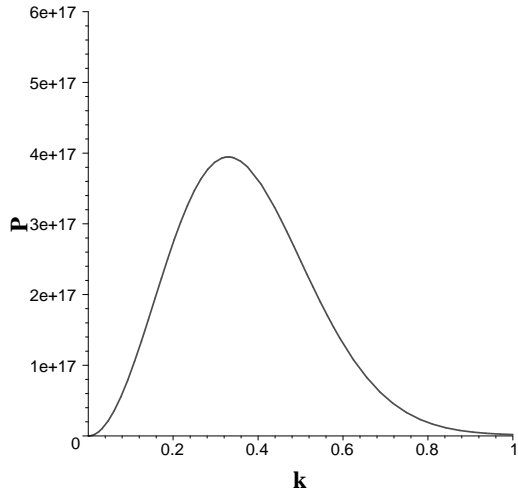}}
\caption{Probability density as a function of $m/\omega$, for $p/p\,'= 0.001$.}
\label{fig:4}
\end{figure}

\begin{figure}[h!t]
\centerline{\includegraphics[scale=0.5]{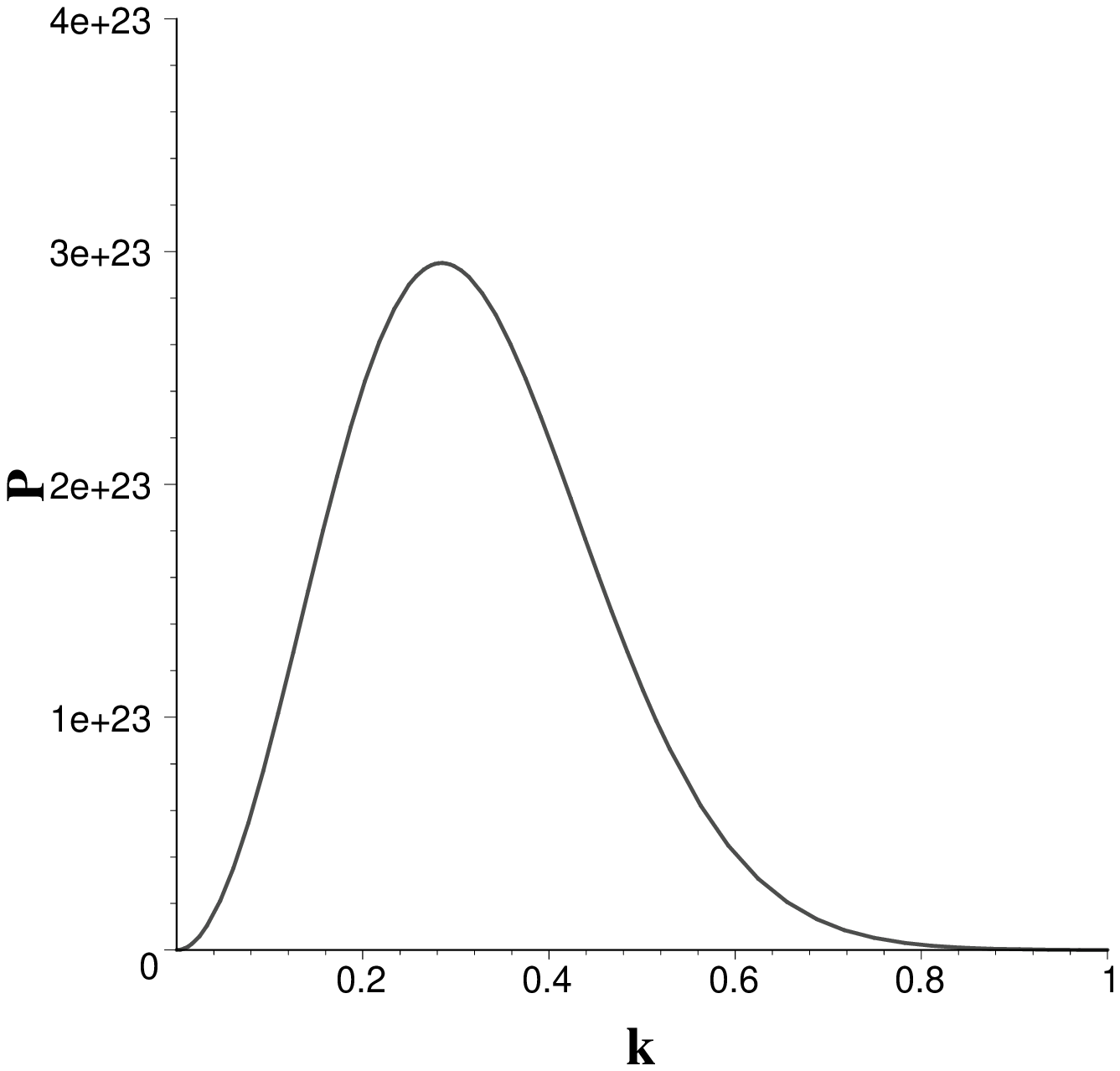}}
\caption{Probability density as a function of $ m/\omega$, for $p/p\,'= 0.0001$.}
\label{fig:5}
\end{figure}

\begin{figure}[h!t]
\centerline{\includegraphics[scale=0.5]{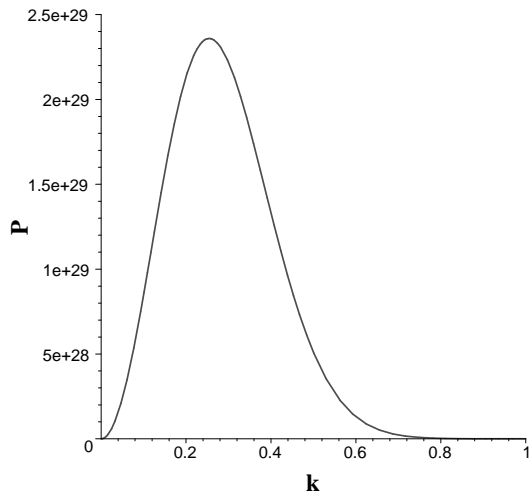}}
\caption{Probability density as a function of $m/\omega$, for $p/p\,'= 0.00001$.}
\label{fig:6}
\end{figure}

\newpage
Our graphs show that, the probability density of pair production  is very important in the limit $\frac{m}{\omega}<1$, which corresponds to the early universe. Also we can observe that, the density of probability is increasing as the ratio $p/p\,'$ is decreasing. This means that it is more probable to produce pairs in which one particle has a small momenta, in comparison with the momenta of the other particle. Our result show that only when the expansion factor is larger than the mass of the particle the probability of pair production is significant. This result show that the probability of particle production was important only in the early universe. Our result is in agreement with Parker's results \cite{B1,B2,B3}, even if we use another method to study the problem of particle production.

The probability density of pair production for the case $\frac{m}{\omega}>\frac{3}{2}$ , is obtained by squaring the amplitude (\ref{re1}):
\begin{eqnarray}
P_{\varphi\varphi^{+}}= \frac{e^{4}}{256\pi^{2}m^{2}}\frac{((\vec{p}-\vec{p}\,')\vec{n})^{2}}{|\vec{p}+\vec{p}\,'|^{2}}\left[\frac{1}{p^{4}}\theta(p-p\,')
\left|f_{\mu}\left(\frac{p\,'}{p}\right)\right|^{2}+\frac{1}{p\,'^{4}}\theta(p\,'-p)\left|f_{\mu}\left(\frac{p}{p\,'}\right)\right|^{2}\right].
\end{eqnarray}
Like in the previous case we want to make a graphical representation of the density of probability as a function of $m/\omega$, for different values of $p/p\,'$. But this time, the numerical values of $m/\omega$ are taken in the interval $[1.5,\, \infty)$, this being the interval where the solutions of the Klein-Gordon equation with imaginary index are correctly defined.
These results are presented in the following graphs:
\begin{figure}[h!t]
\centerline{\includegraphics[scale=0.5]{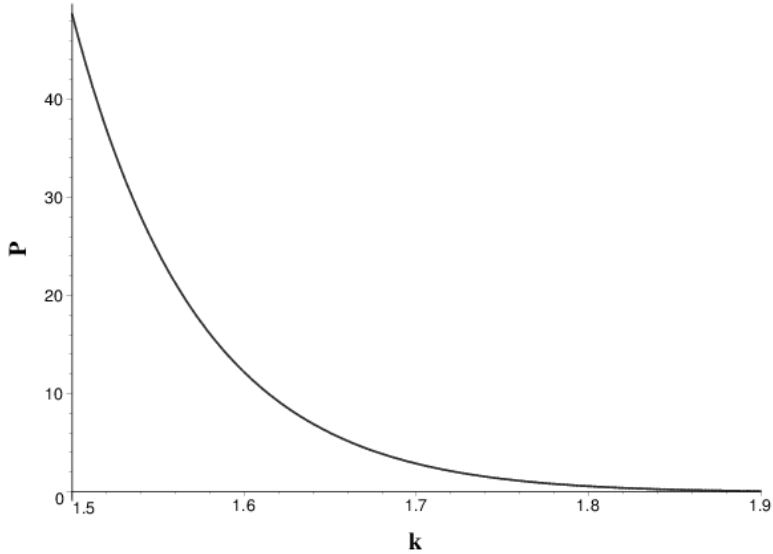}}
\caption{Probability density as function of $m/\omega$, for $p/p\,'= 0.1$.}
\label{fig:1}
\end{figure}
\newpage

\begin{figure}[h!t]
\centerline{\includegraphics[scale=0.5]{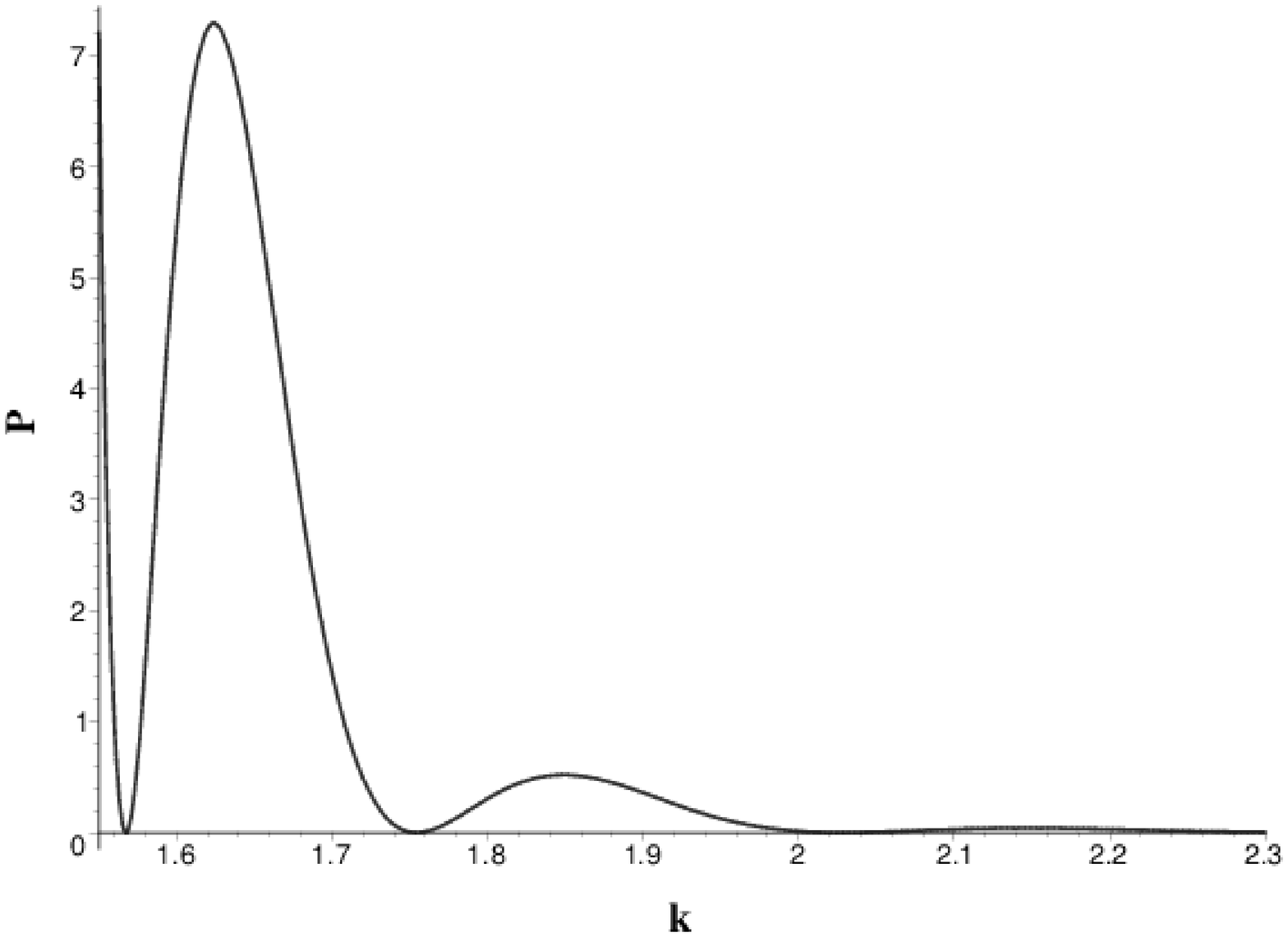}}
\caption{Probability density as a function of $m/\omega$, for $p/p\,'= 0.001$.}
\label{fig:2}
\end{figure}

\begin{figure}[h!t]
\centerline{\includegraphics[scale=0.5]{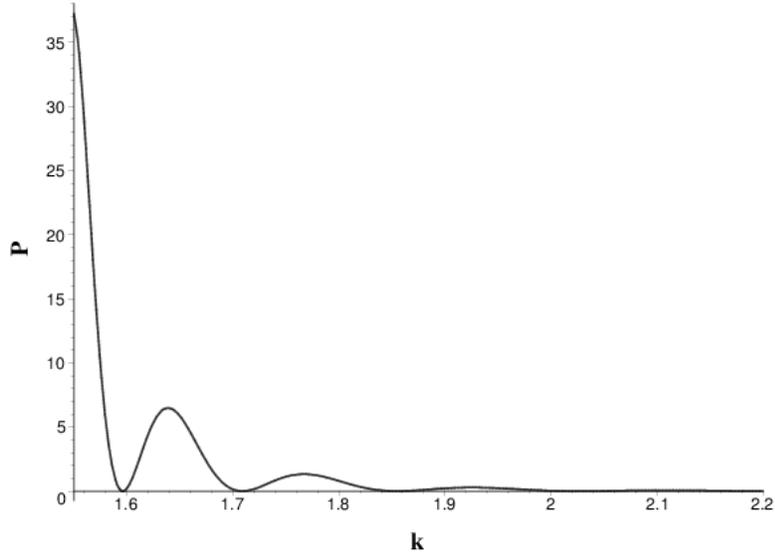}}
\caption{Probability density as a function of $m/\omega$, for $p/p\,'= 0.00001$.}
\label{fig:3}
\end{figure}

From our graphs we see that, the probability density of scalar particle production in the limit $\frac{m}{\omega}>\frac{3}{2}$ is much smaller if we compare with the case $\frac{m}{\omega}<\frac{3}{2}$. The conclusion of this result is the following: the process of scalar particle production from vacuum in the presence of an external electric field, was important only in the early universe. As we can observe from our graphs Figs.(\ref{fig:1})-(\ref{fig:3}) in the Minkowski limit when $\frac{m}{\omega}=\infty$ our probabilities vanish. The Minkowski limit can also be obtained analytically from our results for the amplitude (probability). \\
Further we want to discuss the problem of particle-antiparticle separation.  Our results show that, in both cases the probability densities of pair production are proportional with the term $\frac{[(\vec{p}-\vec{p}\,')\vec{n}]^{2}}{|\vec{p}+\vec{p}\,'|^2}$. Let us consider an arbitrary orientation of momentum vectors, and see what happens with the transition probability density. The spherical coordinates of $\vec{p\,'}$, $\vec{p}$\,, and $\vec{n}$ are: $\vec{p}\,'=(p\,',\alpha,\beta)$, $\vec{p}=(p,\delta,\rho)$, and respectively $\vec{n}=(1,\tau,\sigma)$.
Then we obtain:
\begin{eqnarray}\label{c2}
\frac{[(\vec{p}-\vec{p}\,)\vec{n}]^{2}}{|\vec{p}+\vec{p}\,'|^2}&=& \frac{1}{(p^{2}+p\,'^{2}+2pp\,'\cos\gamma)}\left[(p\,'\sin{\alpha}\cos{\beta}-p\sin{\delta}\cos{\rho})\sin{\tau}\cos{\sigma}\right.\nonumber\\
&&\left.+ (p\,'\sin{\alpha}\sin{\beta}-p\sin{\delta}\sin{\rho})
\sin{\tau}\sin{\sigma}+(p\,'\cos{\alpha}-p\cos{\delta})\cos{\tau}\right].
\end{eqnarray}
Using the expression $(\ref{c2})$ we can consider several particular cases of the orientation of momentum vectors and the unit vector $\vec{n}$, which gives the orientation of the electric field. For example, if $\alpha=\beta=\delta=\rho=\sigma=\tau=\gamma=0$, the probability density of pair production is minimum. This is the case when the two particles are emitted in the direction of the electric field but their momenta are parallel and have the same orientation with electric field vector. The probability density is also minimum if the pair is emitted on the direction of the electric field, but the momenta vectors are parallel and opposite in orientation relatively to the electric field vector. This is the situation when $\alpha=\delta=\pi$, and $\beta=\rho=\sigma=\tau=\gamma=0$. In both cases when the density of probability is minimum, it is more likely that the pair will annihilate in vacuum. The minimum density of probability will be proportional with:
\begin{equation}
\frac{[(\vec{p}-\vec{p}\,')\vec{n}]^{2}}{|\vec{p}+\vec{p}\,'|^2}=\frac{(p-p\,')^{2}}{p^{2}+p\,'^{2}+2pp\,'}
\end{equation}
Contrary to this, we have maximum probability of emission for $\alpha=\beta=\rho=\sigma=\tau=0$ and $\gamma=\delta=\pi$. This is an important case, because it shows how the particle is separated from antiparticle. This is the case when the particles are emitted on the direction of electric field, their momenta are parallel and have opposite orientation. The maximum density of probability will be proportional with:
\begin{equation}
\frac{[(\vec{p}+\vec{p}\,)\vec{n}]^{2}}{|\vec{p}+\vec{p}\,'|^2}=\frac{(p+p\,')^{2}}{p^{2}+p\,'^{2}-2pp\,'}
\end{equation}

Also, is important to  mention that in the case when $\alpha=\beta=\delta=\rho=0$ and $\tau=\sigma=\pi/2$ the probability density of pair production is zero. This means that it is not possible for the pair to be emitted perpendicular on the direction of the electric field.

\section{Conclusions}
We have studied here the scalar particle production, from vacuum in the presence of an electric field on de the Sitter expanding universe. We were able to express the final result of the transition amplitude with the help of the Heaviside step function, gamma-Euler function and the hypergeometric function $_{2}F_{1}$. This result shows that, the momentum is not conserved in this case. Also, we studied the density probability of pair production in the cases $\frac{m}{\omega}>\frac{3}{2}$ and $\frac{m}{\omega}<\frac{3}{2}$. Our graphical representations show that the functions which define the probabilities of pair production are convergent. From our study we found that the pair production of scalar particles form vacuum, in the presence of an external electric field could take place only when the expansion factor was larger than the mass of the particle. This result proves that the massive scalar particles were produced in early universe. Also, in this paper we showed that the probability of pair production vanishes in the Minkowski limit. The form of the probability density of scalar particle production helped us to propose a criterion of separation between particles and antiparticles. This criterion is based on the orientation of the momenta of the two particles. We found that the density of probability is maximum when the particles are moving in opposite direction.

\section{Appendix}

For completing our calculations we use the following integral:
\begin{equation}
\int_{0}^{\infty}dx\,\frac{\sin(ax)}{ax}=\frac{1}{2}sgn(a)\frac{\pi}{a}    \label{a2}
\end{equation}
In the case of temporal integral we use the relation between Hankel function and Macdonald function \cite{B20}:
\begin{equation}
H_{\nu}^{(1,2)}(z)=\mp\,\left(\frac{2i}{\pi}\right)e^{\mp\,i\pi\nu/2}K_{\nu}(\mp\,iz)  \label{a3}
\end{equation}
The form of the resulted temporal integral is \cite{B20}:
\begin{eqnarray}\label{a4}
&&\int_{0}^{\infty}dz z^{-\lambda} K_{\mu}(az)K_{\nu}(bz)=\frac{2^{-2-\lambda}a^{-\nu+\lambda-1}b^{\nu}}{\Gamma\left(1-\lambda\right)}\Gamma\left(\frac{1-\lambda+\mu+\nu}{2}\right)\Gamma\left(\frac{1-\lambda-\mu+\nu}{2}\right)\\   \nonumber
&&\times\Gamma\left(\frac{1-\lambda+\mu-\nu}{2}\right)\Gamma\left(\frac{1-\lambda-\mu-\nu}{2}\right)\,_{2}F_{1}\left(\frac{1-\lambda+\mu+\nu}{2},\frac{1-\lambda-\mu+\nu}{2};1-\lambda;1-\frac{b^{2}}{a^{2}}\right)\\ \nonumber
&&Re(a+b)>0, Re(\lambda)<1-|Re(\mu)|-|Re(\nu)|.
\end{eqnarray}
In our case $\lambda=-1$ and the second condition for convergence
is satisfied. Also in our case $a,b$ are complex
and for solving our integrals we add to $a$ a small real part
$a\rightarrow a+\epsilon$, with $\epsilon>0$ and in the end we
take the limit $\epsilon\rightarrow 0$. This assure the
convergence of our integral and will correctly define the unit
step functions and $f_{\alpha}$ functions.
For expressing our amplitude in final form we use the following relation between hypergeometric functions \cite{B20,B23}:
\begin{eqnarray}\label{a5}
_{2}F_{1}(a,b;c;z)&=& \frac{\Gamma(c)\Gamma(c-a-b)}{\Gamma(c-a)\Gamma(c-b)}\,_{2}F_{1}(a,b;a+b-c+1;1-z)\\ \nonumber
&&+(1-z)^{c-a-b}\,\frac{\Gamma(c)\Gamma(a+b-c)}{\Gamma(a)\Gamma(b)}\,_{2}F_{1}(c-a,c-b;c-a-b+1;1-z).
\end{eqnarray}

\par
\textbf{Acknowledgements}
\par
I would like to thank Professor Ion Cot\u aescu for reading the manuscript and for his observations. I also would like to thank Cosmin Crucean for useful suggestions that help me to improve this work.

This work was supported by the strategic grant POSDRU/159/1.5/S/137750, Project "Doctoral and Postdoctoral programs support for increased competitiveness in Exact Sciences research" cofinanced by European Social Fund within the Sectoral Operational Programme Human Resources Development 2007-2013.


\begin{thebibliography}{99}
\bibitem{B1}
L. Parker, Phys. Rev. Lett. 21, 562 (1968).

\bibitem{B2}
L. Parker, Phys. Rev. 183, 1057 (1969).

\bibitem{B3}
L. Parker, Phys. Rev. D 3, 346 (1971).

\bibitem{B4}
I. I. Cot\u aescu, C. Crucean, Progress of Theor. Phys. 124, 1051 (2010).

\bibitem{B5}
I. I. Cot\u aescu, C. Crucean, A. Pop,  Int. J. Mod. Phys. A 23, 2563 (2008).

\bibitem{B6}
Crucean Cosmin, Phys. Rev. D 85, 084036-1 (2012).

\bibitem{B24}
C. Crucean, Mod. Phys. Lett. A 25, 1679 (2010).

\bibitem{B7}
V. M. Villalba, Phys. Rev. D 60, 127501  (1999).

\bibitem{B8}
E. Schr\" odinger, Physica 6, 899 (1939).

\bibitem{B9}
M. Mijic, Phys. Rev. D 57, 2138 (1998).

\bibitem{B10}
N. D. Birrell, P. C. W. Davies, L.H.Ford, J. Phys. A 13, 961 (1980).

\bibitem{B11}
N. D. Birrell and P. C. W. Davies, Quantum Fields in Curved Space (Cambridge 1982).

\bibitem{B12}
S. Drell and J. D. Bjorken, Relativistic Quantum Fields (New York 1965).

\bibitem{B13}
S. Weinberg, The Quantum Theory of Fields Univ. Press, Cambridge (1995).

\bibitem{B14}
N. A. Chernikov and E. A. Tagirov, Annales Poincare Phys. Theor. A 9, 109 (1968).

\bibitem{B17}
G. V. Shishkin, Class. Quantum Grav. 8, 175 (1991).

\bibitem{B18}
J. Haro, J . Phys. A 44 (2011).

\bibitem{B19}
C. W. Misner, K. S. Thorne and J. A. Wheeler, Gravitation (W. H. Freeman and Company New York, 1973).

\bibitem{B20}
I. S. Gradshtejn and I. M. Ryzhik, Table of Integrals, Series, and Products (Academic Press Inc., San Diego, 1980).

\bibitem{B21}
N. N. Bogoliubov and D. V. Shirkov, Introduction to the theory of quantized fields (Intrscience Publishers, London 1959).

\bibitem{B22}
R. M. Wald, General Relativity (Univ. of Chicago Press, 1984).

\bibitem{B23}
G. N. Watson, Theory of Bessel functions (Cambridge Univ. Press, 1922).



\end{thebibliography}
\end{document}